\documentstyle{amsart}  %
\newsymbol\varkappa 207B
\newcommand{\mathcal}[1]{\cal{#1}}
\newcommand{\mathbb}[1]{\Bbb{#1}}
\newcommand{\mathfrak}[1]{\frak{#1}}

\newcounter{mycomment}
 \newtheorem{lemma}{Lemma}[section]
 \newtheorem{theorem}[lemma]{Theorem}

\theoremstyle{definition}

\newtheorem{conjecture}[lemma]{Conjecture}

\theoremstyle{remark}

\numberwithin{equation}{section}

 \newcommand{\R}{{\mathbb R}}
  \newcommand{\ri}{\rightarrow}

\begin{document}

\title{A   particle in the  Bio-Savart-Laplace magnetic field: explicit solutions}

\author{ D. R. Yafaev} 

\address{Department of Mathematics, University of Rennes-1,\\
 Campus Beaulieu, 35042, Rennes, France}
\email{yafaev@@univ-rennes1.fr}

\maketitle

\begin{abstract}
We consider the Schr\"odinger operator ${\bf H}=(i\nabla+A)^2  $ in the
space $L_2({\mathbb R}^3)$ with a magnetic
 potential $A $  created by an infinite straight current.
We perform a spectral analysis of the operator ${\bf H}$ almost explicitly.
 In particular, we show that the operator ${\bf H}$ is absolutely continuous, its spectrum has
infinite  multiplicity
and coincides with the positive half-axis. Then we find the large-time behavior of solutions
$\exp(-i{\bf H}t)f$ of the time dependent Schr\"odinger equation.
Equations of classical mechanics are also integrated. Our main observation is that 
 both quantum and classical  particles have always a preferable (depending on its charge)
direction of propagation along the current and both of them are confined in the plane orthogonal
to the current.  
\end{abstract}

\section{Introduction}

They are very few examples of explicit solutions of the Schr\"odinger equation with a magnetic
potential.  Probably   the only ones are a constant magnetic field, $B(x,y,z)=B_0$,
(see, e.g., \cite{LL}) and, in the two dimensional case, 
  a   magnetic field localized at the origin, $B(x,y )=B_0\delta(x,y)$ where $\delta(x,y)$ is the
Dirac function  (see 
\cite{AB}).   The solution is expressed in terms of Hermite functions
in the first case and in terms of Bessel functions in the second case.
 Here we suggest a third example of an explicitly solvable   Schr\"odinger equation.
Actually, we consider
the magnetic field $B(x,y,z)$ created by an infinite straight current. Physically,
this case is opposite to the case of an (infinitely) thin
straight solenoid considered in \cite{AB} by Aharonov and Bohm where the field is concentrated
inside  the solenoid.

Suppose that the current coincides with the axis $z$
and that the axes
 $x$, $y$ and $z$ are positively oriented. According to the Bio-Savart-Laplace law (see, e.g.,
\cite{Roc})
\[ 
 B(x,y,z)=\alpha (-r^{-2}y, r^{-2}x,0),\quad r=(x^2+y^2)^{1/2},
\] 
where $|\alpha|$ is proportional to the current strength and $ \alpha>0$ ($ \alpha<0$) if the 
current  streams in the
positive (negative) direction. The magnetic potential is defined by the equation
\[
B(x,y,z)={\rm curl}\, A(x,y,z)
\]
and can be chosen as
\begin{equation}\label{eq:A}  
A(x,y,z)= -\alpha ( 0,0, \ln r).
\end{equation}
Thus, the corresponding Schr\"odinger operator in the space $L_2({\R}^3)$ has the form
\[  
{\bf H}={\bf H}_\gamma=-\partial^2_x  -\partial^2_y + (i\partial_z - \gamma\ln r)^2,
\quad \gamma=e \alpha,
\] 
where $e$ is the charge of a quantum particle of the mass $m=1/2$ and    the speed of the
light $c=1$.

Since   magnetic potential (\ref{eq:A}) grows as $r\ri\infty$, the Hamiltonian ${\bf H}$
does not fit to the well elaborated framework of spectral and scattering theory. Nevertheless, we
perform in Section~2 its spectral analysis almost explicitly. To be more precise, we reduce the
problem to an ordinary differential equation with the potential $\gamma^2 \ln^2 r$  (let us call
it logarithmic oscillator). We show that the operator
${\bf H}$ is
absolutely continuous, its spectrum has infinite multiplicity and coincides with the positive
half-axis. Then we find in Section~3 the large-time behavior of solutions
$\exp(-i{\bf H}t)f$ of the time dependent Schr\"odinger equation.
In Section~4 we integrate equations of classical mechanics. Our main observation is that 
    positively (negatively) charged quantum and classical particles   always move  in the
direction of  the current (in the opposite direction) and are localized in the orthogonal plane.  

A detailed presentation of the results of this note can be found in \cite{Y}.

\section{Spectral analysis of the operator ${\bf H}$}

Let us first consider   a more general magnetic potential
 \begin{equation}\label{eq:Ag}  
A(x,y,z)=   ( 0,0, {\cal A}(x,y))
\end{equation}
with an arbitrary (we disregard here domain questions) real function ${\cal A}( x,y)$ which tends
to infinity (either
$+\infty$ or
$-\infty$) as
$r=(x^2 +y^2)^{1/2}\ri\infty$. The corresponding Schr\"odinger operator is
\[  
{\bf H}=- \Delta + (i\partial_z + e {\cal A}(x,y))^2,
\]  
where $\Delta$ is always the Laplacian in the variables $(x,y)$.
 Since ${\cal A}$ does not depend on $z$, we make the Fourier
transform
$\Phi=\Phi_z$ in the variable
$z$. Then the operator
$H=\Phi {\bf H}\Phi ^*$ acts in the space $L_2({\R}^2\times {\R})$ as 
\[
 (Hu)(x,y,p)= (h(p)u)(x,y,p),
\] 
 where
\begin{equation}\label{eq:shhn}  
   h(p) =- \Delta+ (p-e {\cal A}( x,y))^2.
\end{equation}
Here $p\in\R$ (the momentum in the
direction of the $z$-axis) is the   variable dual to $z$
and  the operator $ h(p)$ acts in the space $L_2({\R}^2)$.
 
 Since ${\cal A}(x,y)\ri\infty$ or ${\cal A}(x,y)\ri-\infty$  as
$r\ri\infty$, the spectrum of each operator
$h(p) $ is positive and discrete. Let $\lambda_n(p)$,
$n=1,2,\ldots$, be its eigenvalues, numerated in such a way that 
$\lambda_n(p)$ are analytic functions of $p$. The spectrum of the
operator
$H$, and hence of
${\bf H}$, consists of the sets (branches) covered by the functions  $\lambda_n(p)$,
$n=1,2,\ldots$,  as $p$ runs from $-\infty$ to  $\infty$. This is similar both to the cases of the
constant field where $\lambda_n(p)=|e B_0| (2n+1 )+p^2$ and to the periodic problem where the role
of the momentum $p$ in the direction of the $z$-axis is played by the quasimomentum (see, e.g.
\cite{RS}). Thus, the general Floquet theory implies that the spectrum of $H$   is absolutely
continuous, up  eventually  to some eigenvalues of infinite multiplicity. Such eigenvalues  
appear if at least one of the functions
$\lambda_n(p) $ is a constant  on some interval. Then this function is a constant  for all
$p\in\R$.
On the other hand, if, say, the function
$-e {\cal A}( x,y)$ is semibounded from below, then
\begin{equation}\label{eq:prel}
\lim_{p\rightarrow \infty}\inf_{( x,y)\in{\R}^2}(p-e {\cal A}( x,y))^2=\infty 
\end{equation} 
 as $p\rightarrow \infty$, and hence $\lim_{p\rightarrow \infty}\lambda_n(p)=\infty$ for all
$n$. 
Thus, we have the following simple result.

\begin{theorem}\label{prel} 
Suppose that $ {\cal A}(x,y)$ is a semi-bounded function which tends either to $+\infty$
or to $-\infty$ as $ r\ri\infty$.
Then the   operator    ${\bf H}$ is absolutely continuous.
\end{theorem}

Note that the Thomas arguments (see, e.g., \cite{RS}) relying on the study  
of the operator-function
$h(p)$ for complex $p$ are not necessary here.

\medskip

The problem may be further simplified if $ {\cal A}(x,y)={\cal A}(r)$. Then
 we can separate variables in the polar coordinates $(r,\theta)$. Denote by ${\cal H}_m$
the space of functions $u(r)e^{im\theta}$ where $u \in L_2({\R};rdr)$  and  $m=0,\pm 1,\pm
2,\ldots$ is the orbital quantum number. Then
\begin{equation}\label{eq:HHm}  
    L_2({\R}^2)=\bigoplus_{m=-\infty}^\infty {\cal H}_m.
\end{equation}
Every subspace ${\cal H}_m$ is invariant with respect to the operator $ h(p)$.
The spectra of their restrictions $ h_m(p)$ on  ${\cal H}_m$  consist  of positive simple
eigenvalues
$\lambda_{m,1}(p)< 
\lambda_{m,2}(p)<\ldots$,  which are analytic functions of $p$. We denote by $\psi_{m,1}(r,p),
\psi_{m,2}(r,p),\ldots$   the corresponding eigenfunctions which  are supposed to be normalized and
real. 

\medskip

Let us return to the operator ${\bf H}$ with the potential  $ {\cal A}( r)=-\alpha \ln r$.
 In this case operator (\ref{eq:shhn} ) equals 
\[
   h(p) =- \Delta+ \ln^2(e^p r^\gamma).
\]
Since
$ \overline{ {\bf H}_\gamma   u  }= {\bf H}_{-\gamma}  \overline{  u }$,
  it suffices to consider the case $\gamma>0$.
It is convenient to transfer the dependence on the momentum $p$
into the kinetic energy and to introduce the parameter $ a= 
e^{p/ \gamma}\in (0,\infty)$ instead of
$p$. Let us set
\begin{equation}\label{eq:sht}  
   K (a) = -a^2 \Delta+ \gamma^2 \ln^2 r,
\end{equation}
 and let $w(a)$,  
$ (w(a)f(x,y) = a f(a x,ay)$,
 be the unitary operator of dilations in the space $L_2({\R}^2)$. Then
\begin{equation}\label{eq:wh}  
    w^*(a) h (p)w(a) = K (a),\quad a=  e^{p/\gamma}.
\end{equation}
 We denote by $\mu_{m, n}(a)$ and $\phi_{m,n}(r,a)$
eigenvalues and eigenfunctions of the restrictions of the
  operators $K(a) $ on the subspaces ${\cal H}_m$. It follows from (\ref{eq:wh}) that $\mu_{m,
n}(a)=\lambda_{m, n}(p)$ and
$\phi_{m, n}(a)=w^*(a)\psi_{m, n}(p)$. 
Actually, decomposition (\ref{eq:HHm}) is needed only to avoid crossings between different
eigenvalues of the operators $h(p)$. It allows us to use   formulas of perturbation theory
(see, e.g., \cite{RS}) for simple eigenvalues.  We   fix $m$ and omit it from the notation.

The next assertion  is quite elementary but plays the crucial role in the following. 

\begin{lemma}\label{ev}
 For every $n$, we have that $ \mu^\prime_n(a)>0$ for all $a>0$.
\end{lemma}
 
Indeed,  analytic perturbation theory shows   that
\begin{equation}\label{eq:PT} 
\mu^\prime_n(a)=(K_n^\prime (a)\phi _n( a), \phi _n( a))
 =2a\int_{{\R}^2}  |\nabla\phi _n(x,y,a)|^2dxdy.
\end{equation} 
This expression is obviously   positive since otherwise
$\phi_n(x,y,a)={\rm const}$.

The next assertion realizes an obvious idea that  the spectrum of $K(a)$
converges  as $a\rightarrow 0$ (in the  quasiclassical limit) to that of the multiplication operator
by
$\gamma^2\ln^2 r$, which is continuous and starts from zero. 

 \begin{lemma}\label{ev1} For every $n$, we have that  
$\lim_{a\rightarrow 0}\mu_n(a)=0$.
\end{lemma}
 
Since the function $\ln r$ is not semibounded, relation (\ref{eq:prel})
is not true in our case. Nevertheless,   taking
into account the kinetic energy, we obtain the following result.

\begin{lemma}\label{ev2} For every $n$, we have that  
$\lim_{a\rightarrow \infty}\mu_n(a)=\infty$.
\end{lemma}

In terms of eigenvalues $\lambda_n(p)$ of the operators $h(p)$, Lemmas~\ref{ev} --
\ref{ev2}  mean  that $  \lambda_n^{\prime }(p)>0$ for all $p\in\R$ and
$\lim_{p\rightarrow - \infty}\lambda_n (p)=0$, 
 $\lim_{p\rightarrow  \infty}\lambda_n (p) =\infty$ (for $\gamma>0$).

Let $\Lambda_n$ be multiplication operator by the function $\lambda_n(p) $ in
the space $L_2({\R})$. It follows from the results on the function $\lambda_n(p) $ that the
spectrum of $\Lambda_n$ is absolutely continuous, simple and coincides with the positive half
axis. Let us introduce a  unitary mapping  
\[
  \Psi:L_2({\R}_+\times {\R};rdrdp)\rightarrow\bigoplus_{n=1}^\infty L_2({\R})
\] 
by the formula
$   (\Psi f)_n(p)=\int_0^\infty f(r,p)  \overline{\psi_n(r,p)} r dr$.
Then
\begin{equation}\label{eq:W2}  
     \Psi\Phi {\bf H}\Phi^* \Psi^*=\bigoplus_{n=1}^\infty \Lambda_n 
\end{equation}
(of course ${\bf H}={\bf H}_m$ and $\Lambda_n=\Lambda_{n,m}$), and we obtain the following

\begin{theorem}\label{sp} 
The spectra of all operators ${\bf H}_m$ and ${\bf H}$ are absolutely continuous, have infinite
multiplicity  and coincide with the positive half axis.
\end{theorem}

As a by-product of our considerations, we have constructed a complete set of eigenfunctions of
the operator ${\bf H}$. They are parametrized by the orbital quantum number $m$,
the momentum $p$ in the direction of the $z$-axis and the number $n$ of
an eigenvalue $\lambda_{m,n}(p)$ of the operator $h_m(p)$ defined by formula (\ref{eq:shhn})
on the subspace ${\cal H}_m$. Thus, if we
set
\[
{\bf u}_{m,n,p}(r,z,\theta)=e^{ipz} e^{im\theta} \psi_{m,n}(r,p),
\]
then
${\bf H}{\bf u}_{m,n,p}=\lambda_{m,n}(p) {\bf u}_{m,n,p}$.

\section{Time evolution}

Explicit formulas obtained in the previous section allow us to find the asymptotics for large $t$
of solutions $u(t)=\exp(-i{\bf H} t)u_0$ of the time dependent Schr\"odinger equation. 
On every subspace with a fixed
orbital quantum number $m$, the problem reduces to the asymptotics of the function 
 $u(t)=\exp(-i {\bf H}_m t)u_0$.  Below we fix $m$ and suppose that $ \gamma>0$.

 Assume that 
\begin{equation}\label{eq:InV}
(\Phi u_0)(r,p)=\psi_n(r,p)f(p),
\end{equation}
where $f\in C_0^\infty({\R})$. Then it follows from formula (\ref{eq:W2}) that
\begin{equation}\label{eq:TE}  
     u(r,z,t)=(2\pi)^{-1/2}\int_{-\infty}^\infty e^{ipz-i\lambda_n(p) t}\psi_n(r,p)f(p)dp.
\end{equation}
The stationary points of this integral are determined by the equation
\begin{equation}\label{eq:stp}  
     z=\lambda_n^\prime(p)t.
\end{equation}
Since $  \lambda_n^\prime(p)>0$, the equation (\ref{eq:stp}) has a solution
only if $  zt >0$. 
 We need the following information on the eigenvalues $\mu_n(a)$ of the operator
(\ref{eq:sht}).

\begin{lemma}\label{eigenval}  For every $n$, we have that  
$
\lim_{a\rightarrow 0}a \mu^\prime_n(a)=0.
$
\end{lemma}

Indeed, it follows from equation (\ref{eq:PT}) that
$
a\mu_n^\prime(a)\leq 2 \mu_n(a).
$
Therefore it remains to use Lemma~\ref{ev1}. 

Lemma~\ref{eigenval} means that $\lim_{p\rightarrow - \infty}\lambda_n^\prime(p)=0$.
The following conjecture  is  physically quite plausible and is used mainly to formulate
Theorem~\ref{time} below in a simpler form.

\begin{conjecture}\label{eigenvalues} 
For every $n$, we have that $\lambda_n^{\prime\prime}(p)>0$ for all $p\in\R$ and
$\lim_{p\rightarrow  \infty}\lambda_n^\prime(p) =\infty$.
\end{conjecture}
 
 Therefore equation
$\lambda_n^\prime(p)=v$
has a unique solution $p_n=\varphi_n(v)$ for every $v>0$. Clearly,
\begin{equation}\label{eq:tech}  
     \lambda_n^{\prime\prime}(\varphi_n(v))\varphi_n^\prime(v)=1.
\end{equation}
Let   
      $\Phi_n(v)=\varphi_n(v)\alpha-\lambda_n(\varphi_n(v))$,
  $\theta(v)=1$ for $v>0$, $\theta(v)=0$ for $v <0$ and
 $\pm i=e^{\pm \pi i/2}$.
Applying to the integral (\ref{eq:TE}) the stationary phase method and taking into account
identity (\ref{eq:tech}), we find that   
\begin{equation}\label{eq:TE1}  
     u(r,z,t)=e^{i\Phi_n(z/t)t }
\psi_n(r,\varphi_n(z/t))\varphi_n^\prime(z/t)^{1/2}f(\varphi_n(z/t))(it)^{-1/2}\theta (z/t)
+ u_\infty(r,z,t),
\end{equation}
where
\begin{equation}\label{eq:TE1r} 
\lim_{t\rightarrow\pm\infty }|| u_\infty(\cdot,t)||= 0.
\end{equation}
 Note that the norm in the space $L_2({\R}_+\times{\R})$
of the first term in the right-hand side of (\ref{eq:TE1}) equals $||u_0||$.
The asymptotics (\ref{eq:TE1}) extends of course to all $f\in L_2({\R})$
and to linear combinations of functions (\ref{eq:InV}) over different $n$.
 Thus, we have proven

\begin{theorem}\label{time}
Assume that Conjecture~$\ref{eigenvalues}$ is fulfilled. 
Suppose that   $\gamma>0$. 
Let  $u(t)=\exp(-i {\bf H}_m t) u_0$ where $u_0$ satisfies $(\ref{eq:InV})$. Then the
asymptotics as $t\rightarrow \pm \infty$ of this function  
 is given by relations $(\ref{eq:TE1})$,
$(\ref{eq:TE1r})$. Moreover, if $f\in C_0^\infty({\R})$   and $\mp z
>0$, then the function $ u(r,z,t)$ tends to zero faster than any power of $|t|^{-1}$
as $t\rightarrow \pm \infty$.

Conversely, for any $g\in L_2({\R}_+)$ define the function $u_0$ by the equation
\[
(\Phi u_0)(r,p)=\psi_n(r,\lambda_n^\prime(p))\lambda_n^{\prime\prime}(p)^{1/2}
g(\lambda_n^\prime(p)).
\]
Then $u(t)=\exp(-i {\bf H}_m t) u_0$ has the asymptotics as $t\rightarrow \pm \infty$ 
\[
     u(r,z,t)=e^{i\Phi_n(z/t)t }
\psi_n(r, z/t )g( z/t)(it)^{-1/2}\theta (z/t) +
u_\infty(r,z,t),
\]
where $u_\infty$ satisfies $(\ref{eq:TE1r})$.
\end{theorem}

\section{Classical mechanics}

 Let us consider
the motion of a classical particle of mass $m=1/2 $ and charge $e$ in a magnetic field
created by potential (\ref{eq:Ag}) where ${\cal A} (x,y)={\cal A} (r)$, $r=(x^2+y^2)^{1/2}$.
We suppose that  ${\cal A} (r)$ is an arbitrary $C^2 $-function such that ${\cal A}
(r)=o(r^{-1})$ as
$r\rightarrow 0$ and $|{\cal A} (r)|\rightarrow\infty$ as
$r\rightarrow \infty$. The solution given below is, to a large extent, similar to the Kepler
solution of equations of motion for a particle in a spherically  symmetric electric field.
However, in the electric case the motion is always restricted to a plane, whereas in the magnetic
case it is confined in the plane $z=0$ but the propagation of a particle in the $z$-direction has
a non-trivial character. We proceed here from the Hamiltonian formulation.  An approach based on 
the Newton equations can be found in \cite{Y}.

Let ${\bf r}$  be a position of a particle and   ${\bf p}$ be its momentum.
Let us write down the Hamiltonian
 \[
H({\bf r}, {\bf p})= ({\bf p}^2-e A({\bf r}))^2
\] 
in the cylindrical coordinates $(r, \varphi, z)$. In the case (\ref{eq:Ag}) where ${\cal A}
(x,y)={\cal A} (r)$, we have that 
\[
 H({\bf r}, {\bf p})= (p_r^2+r^{-2}p_\varphi^2)
+  (p_z -e {\cal A} (r))^2,
\] 
where $p_r$, $p_\varphi$ and $p_z$ are momenta conjugated to the coordinates
$ r$, $ \varphi$ and $ z$. Since $ H({\bf r}, {\bf p})$ does not depend on 
$ \varphi$ and $ z $, the momenta $p_\varphi(t)$ and $p_z(t)$ are
conserved, i.e., $p_\varphi(t)=M$ ($M$ is the moment of momentum with respect to the $z$-axis)
and 
$p_z(t)=P$ (the magnetic momentum in the $z$-direction). Therefore Hamiltonian equations read as
\begin{equation}\label{eq:hamr}\left.\begin{array}{lcl} 
r^\prime(t)&=&2 p_r(t),
\\
p_r^\prime(t)&=&-V^\prime(r(t)),
\end{array}\right\}
\end{equation}
where
\begin{equation}\label{eq:V}
V(r)=M^2 r^{-2}  
+ (P  -e {\cal A} (r))^2,
\end{equation}
and
\begin{equation}\label{eq:ham2}
\varphi^\prime(t)= 2 M  r(t)^{-2}
\end{equation}
\begin{equation}\label{eq:ham3}
z^\prime(t)=   2(P  -e {\cal A} (r(t))).
\end{equation} 

It suffices to solve the system (\ref{eq:hamr}) since, given $r(t)$, the solutions of equations
(\ref{eq:ham2}) and (\ref{eq:ham3}) are constructed by the formulas
\begin{equation}\label{eq:theta} 
\varphi (t)     =\varphi (0)+ 2 M\int_0^t r(s)^{-2}ds 
\end{equation}
and
\begin{equation}\label{eq:NEzpint1}
  z (t) -z (0)= 2  \int_0 ^t   (P-e  {\cal A}(r(s)) ) d s.
\end{equation}
The solution of the system (\ref{eq:hamr}) is quite similar to the solution of
the  Kepler problem although in our case the effective potential energy
(\ref{eq:V}) depends additionally on the momentum $P$ in the $z$-direction.
In the solutions of the quantum problems, it is reflected by the fact that, for electric
spherically symmetric potentials, the variables can be separated
(in the spherical coordinates), whereas in our case the operators $h(p)$ depend on $p$.

Thus, to solve (\ref{eq:hamr}), we remark that
\begin{equation}\label{eq:Kpr}
 4^{-1}   r^{\prime}(t)^2+ V(r (t)) =K,
\end{equation}
where $K= 4^{-1}  r^\prime(0)^2 + M^2 r(0)^{-2}  + 4^{-1}  z^\prime(0)^2 $ is a
constant kinetic energy of a particle.  Clearly, (\ref{eq:Kpr}) is the equation of
one-dimensional motion  (see, e.g., \cite{LLcl}) with the
effective potential  energy $ V(r)$ and the total energy $K$. It admits the separation of
variables and can be integrated by the formula
\begin{equation}\label{eq:Int}
  t=\pm 4\int\Bigl(K -V(r)\Bigr)^{-1/2} dr.
\end{equation}
Note that
$ V(r)\rightarrow\infty$ as $r\rightarrow 0$ and $r\rightarrow \infty$. 
 Let $r_{\min}$ and
$r_{\max}$ be the roots of the equation $V(r)=K $ 
($r_{\min}$ and $r_{\max}$ are the nearest to $r(0)$ roots such that
$r_{\min}\leq r(0) \leq r_{\max}$). It follows from
(\ref{eq:Int}) that the function $r(t)$ is periodic with period
\begin{equation}\label{eq:IntT}
  T=8\int_{r_{\min}}^{r_{\max}}\Bigl(K -V(r)\Bigr)^{-1/2} dr 
\end{equation}
and $r_{\min}\leq r(t) \leq r_{\max}$.
One can imagine, for example, that on the period   the function $r(t)$ increases
monotonically from 
$r_{\min}$ to $r_{\max}$ and then decreases from  $r_{\max}$ to $r_{\min}$.
Thus, we have integrated the system (\ref{eq:hamr}) and (\ref{eq:ham2}), (\ref{eq:ham3}).

\begin{theorem}\label{cl} 
 In the variable $r$ a classical particle moves
periodically according to   equation $(\ref{eq:Int})$ with period $(\ref{eq:IntT})$. The angular
variable is determined by equation $(\ref{eq:theta})$ so that $\varphi(t)$ is a monotone
function of $t$ and 
$\varphi(t) = \varphi_0 t + O(1)$, where
$\varphi_0= 2 M T^{-1} \int_0^T r(s)^{-2}ds$,
 as $|t|\rightarrow \infty$. 
 The  variable $z(t)$ is determined by equation $(\ref{eq:NEzpint1})$.
\end{theorem}

According to equation (\ref{eq:ham3}) a particle can move  in the direction of the current
as well as in the opposite direction.
Nevertheless one can give simple sufficient conditions for the inequality 
\begin{equation}\label{eq:ZTz}
\pm (z(t+T) - z(t))> 0 
\end{equation}
(for all $t$). Indeed, it follows from the Newton equation $ r^{\prime\prime}(t)=-2V^{ \prime}(r(t))$ (which is
a consequence of (\ref{eq:hamr})) and expression  (\ref{eq:V}) that
\[
  r^{\prime\prime}(t)= 4M^2 r^{-3}(t) + 4 e {\cal A}^\prime (r(t)) (P-e {\cal A}  (r(t))).
\]
Using equation (\ref{eq:ham3}), we see that
\[
2e z^{ \prime}(t)=
(  r^{\prime\prime}(t)-4M^2 r^{-3}(t) )
  {\cal A}^\prime (r(t))^{-1}.
\]
Integrating this equation   and taking into account   periodicity of the
function $r(t)$, we see that, for all $t$,
\begin{eqnarray}\label{eq:ZT}
2 e ( z(t+T) -z (t) )= 
  \int_0^{ T}  r^{\prime\prime}(s) {\cal A}^\prime (r(s))^{-1}ds
- 4 M^2\int_0^{ T}  r (s)^{-3}{\cal A}^\prime (r(s))^{-1} ds
\nonumber\\
=   \int_0^{ T}  r^{\prime}(t)^2 {\cal A}^\prime (r(t))^{-2}{\cal A}^{\prime\prime} (r(t))dt -
4 M^2\int_0^{ T}  r (s)^{-3}{\cal A}^\prime (r(s))^{-1} ds.
\end{eqnarray}
Let us formulate the results obtained.

\begin{theorem}\label{up} 
 The increment of the
variable $z$ on every period is determined by equation
$(\ref{eq:ZT})$. In particular, if $\pm e  {\cal A}^\prime (r ) < 0$ and
$\pm e  {\cal A}^{\prime\prime} (r )\geq 0$ for all $r$, then inequality $(\ref{eq:ZTz})$ holds.
In this case $z(t)= z_0 t + O(1)$ with $ z_0=T^{-1}(z(T) -z (0))  $, $\pm z_0>0$, as
$|t|\rightarrow \infty$.
\end{theorem}

In particular, for potentials ${\cal A}  (r )=-\alpha \ln r$ and
${\cal A}  (r )=-\alpha  r^a$ where $a\in (0,1)$, inequality  (\ref{eq:ZTz})  holds
if $\pm e \alpha>0$. Note that in these cases the fields $B(x,y,z)=  {\cal A}^\prime  (r )
r^{-1}(y,-x,0)$tend to $0$ as $r\ri\infty$.

It follows from equation (\ref{eq:ham3}) that if, say, 
$ e  {\cal A}^\prime (r ) < 0$ and the
point $r_{\rm cr}$ is determined by the equation 
$p=e {\cal A}  (r_{\rm cr} )$, then $z(t)$ increases for $r(t)\in (r_{\rm cr},r_{\max})$ and
decreases for $r(t)\in (  r_{\min},r_{\rm cr})$.
Of course, it is possible that  $ r_{\rm cr}<r_{\min} $. In this case, $z(t)$ always increases.
Let us discuss this phenomena in more details on our leading example 
${\cal A}  (r )=-\alpha \ln r$.   Then
$r_{\rm cr}=e^{-P/ \gamma}$ where $\gamma=e\alpha$. The points $r_{\min}$ and $r_{\max}$ are
determined from the equation 
\[
V(r)=M^2r^{-2}+\ln^2(e^P r^\gamma)=K.
\]
The function $z(t)$ is increasing for all $t$ if $r_{\rm cr}<r_{\min} $ or, equivalently,
$V(r_{\rm cr})\geq K$ and 
$ r_{\rm cr}\leq r(0)$. The first of these conditions is equivalent to
$M^2 e^{2P/ \gamma}\geq K$ or, since in view of (\ref{eq:ham3}) 
$e^{2P/ \gamma}=r(0)^{-2} e^{z^\prime(0)/ \gamma}$, to
\[
M^2 r(0)^{-2}  e^{z^\prime(0)/ \gamma}\geq
4^{-1}r^\prime(0)^2 + M^2 r(0)^{-2}  + 4^{-1} z^\prime(0)^2.
\]
Thus, $z^\prime(0)$ should be a sufficiently large positive number ($z^\prime(0)\leq 0$ is
definitely excluded). In this case the condition $ r_{\rm cr} \leq r(0)$ which is equivalent to
$z^\prime(0)\geq 0$ is automatically satisfied. Note finally that always $r_{\rm cr}\leq r_{\max}
$, that is the function $z(t)$ cannot be everywhere decreasing (this is of course also a
consequence of Theorem~\ref{up}). Indeed,  inequality $r_{\rm cr}\geq r_{\max} $ is
equivalent to
$V(r_{\rm cr})\geq K$ and  $ r_{\rm cr} \geq r(0)$. The first of them require that 
$z^\prime(0)> 0$ while the second require that 
$z^\prime(0)\leq 0$.
 
 Thus,      positively (negatively)
charged classical and quantum particles always move asymptotically   in the direction  of the
current (in the opposite direction).   In the
plane orthogonal to the direction of the current classical and quantum particles are essentially
localized.

\end{document}